\documentstyle[12pt,epsf,epsfig,wrapfig,amsfonts,amssymb]{article}
\newcommand{\be}{\begin{eqnarray}}
 \newcommand{\ee}{\end{eqnarray}}
 \newcommand{\nee}{\nonumber\end{eqnarray}}

\begin{document}

\begin{center}
{\bfseries ABOUT $s-\bar s$, $\Delta s-\Delta\bar s$ AND $D_d^{K^+-K^-}$
  IN $K^\pm$ PRODUCTION IN SIDIS }
\vskip 5mm
\underline{E. Christova}$^{1\dag}$ and E. Leader$^{2 }$

\vskip 5mm
{\small
(1) {\it
Institute for Nuclear research and Nuclear Energy, Bulgarian Academy of Sciences, Sofia
}\\
(2) {\it
Imperial College, London
}\\
$\dag$ {\it
E-mail: echristo@inrne.bas.bg
}}
\end{center}

\vskip 5mm
\begin{abstract}
We consider semi-inclusive unpolarized DIS for the
production of charged Kaons and the different possibilities, both
in LO and NLO, to test the conventionally used assumptions $s-\bar
s=0$, $\Delta s-\Delta \bar s=0$ and $D_d^{K^+-K^-}=0$. The considered tests have the
advantage that they do not require any knowledge of the fragmentation functions.
\end{abstract}

\vskip 8mm

{\bf 1. Introduction}
\vspace{0.5cm}

Inclusive deep inelastic
scattering (DIS) gives information about the parton densities (PD)
$q+\bar q$ and $\Delta q+\Delta \bar q$. Analogously,
 $e^+e^-\rightarrow hX $ gives information about the fragmentation functions (FF) $D_q^{h+\bar h}$.
 However, the new generation of
 semi-inclusive DIS (SIDIS) experiments  performed with increasing precision and variety
 during the last years,
  present a new  powerful instrument to reveal in more details the
spin and flavour structure of the nucleon. However, as data is still
not enough and  not precise enough,
it has become  conventional to make certain reasonable sounding
assumptions in analyzing the data. The usually made assumptions are:
\be
s(x)=\bar s(x),\quad \Delta s(x) =\Delta \bar s(x),\quad D_d^{K^+}(z) =D_d^{K^-}(z).
\ee
In this paper we discuss to what extent these assumptions can be
justified and tested experimentally, in both, LO and NLO in QCD.  We
suggest possible tests for the reliability of the leading order (LO)
treatment of the considered processes. The considered tests do not
 require any knowledge of the  (FFs). In more details these results are published
 in \cite{kaons_we}.
\vspace{0.5cm}

%%%%%%%%%%%%%%%%%%%%%%%%%%%%%%%%%%%%%%%%%%%%%%%%%%%%%%%%%%%%%%%%%%%%%%%%%%%%%%%%%%%%%%%
{\bf 2. Positivity constraints}
%%%%%%%%%%%%%%%%%%%%%%%%%%%%%%%%%%%%%%%%%%%%%%%%%%%%%%%%%%%%%%%%%%%%%%%%%%%%%%%%%%%%%%%
\vspace{0.5cm}

Here we discuss what we can learn about the strange quark densities
 from positivity conditions.
If $s_+ \,(\bar s_+)$ and $s_- \,(\bar s_-)$ denote the $s (\bar s)$-quarks with
helicities along and opposite the helicity of the nucleon,
 the unpolarized and polarized parton densities are defined as follows:
 \be
 s=s_++s_-,\quad \bar s=\bar s_++\bar s_-,\quad
\Delta s=s_+-s_-,\quad \Delta\bar s=\bar s_+-\bar s_-.
 \ee
 Then from $s_{\pm}\geq 0$ and $ \bar s_{\pm}\geq 0$ the following
positivity constraints follow:
\be
\vert  s- \bar s\vert \leq
s+\bar s ,\quad \vert  \Delta s\pm\Delta \bar s\vert \leq s+\bar
s. \label{posit}
\ee
i.e. all parton densities are constrained only by
$s+\bar s$, our knowledge of the sum $\Delta s+\Delta \bar
s$ does not put any additional limits.
Note that $s- \bar s \lessgtr 0$ and $ \Delta s\pm\Delta \bar
s \lessgtr 0$.

From experiment, it
is known with a good accuracy  that $s+\bar s$ is different from zero
only for small $x\lesssim  0.4$. Then (\ref{posit}) implies that only in this same interval,
 $x\lesssim 0.4$, the combinations
$ s-\bar s$ and $\Delta s\pm \Delta\bar s$ can be different from zero. Also,
as $\int_0^1 dx (s-\bar s)=0$,  it follows that $(s-\bar s)$  changes sign in $x=[0,0.4]$.
\vspace{0.5cm}

{\bf 3.  SIDIS $e+N\to e+K^\pm+X$ }
\vspace{0.5cm}

 Further we shall work with the difference cross sections in SIDIS.
As shown in~\cite{Dubna05}, the general expression for $K^\pm$ production in SIDIS  is:
 \be
 \quad\tilde \sigma_p^{K^+-K^-}(x,z)=\frac{1}{9}\left[4  u_V\otimes
D_u +  d_V\otimes D_d + (s -\bar s)\otimes
D_s\right]^{K^+-K^-}\otimes  \hat\sigma_{qq} (\gamma q \to q
X)\label{diffp}\\
 \quad \tilde \sigma_n^{K^+-K^-}(x,z) =\frac{1}{9}\left[4
d_V\otimes D_u +  u_V\otimes D_d + (s -\bar
s)\otimes D_s\right]^{K^+-K^-}\otimes  \hat\sigma_{qq} (\gamma q
\to q X).\label{diffn}
 \ee
 Here $D_q^{K^+-K^-} \equiv D_q^{K^+}-D_q^{K^-}$,
$\hat\sigma_{qq}$ is  the perturbatively  calculable, hard
partonic cross section $q\gamma^*\to q+X$:
\be
 \hat\sigma_{qq} &=&  \hat\sigma_{qq}^{(0)} +
 \frac{\alpha_s}{2\pi} \hat\sigma_{qq}^{(1)}\,,
 \ee
normalized so that the LO contribution is $ \hat\sigma_{qq}^{(0)} = 1 $.
 For simplicity, we use  $\tilde\sigma_N^{K^\pm}$ and
$\tilde\sigma_N^{DIS}$ in which common kinematic factors have been
removed~\cite{strategy}.

As shown in~\cite{strategy}, the advantage of the difference cross sections
is that all terms in $\sigma_N^{K^+-K^-}$ are non-singlets both in PD and FF.
 This implies that 1) gluons do not enter -- neither $g(x)$ nor $D_g^h(z) $ -- and 2) their
 $Q^2$-evolution is rather simple.

  As $D_s^{K^+-K^-}$ is a favoured transition
and thus expected to be big,  eqs. (\ref{diffp}) and (\ref{diffn}) show that
$\sigma_N^{K^+-K^-}$
are  sensitive  to the  combination $(s-\bar s)$ which we are interested in.
Up to now all analyses of  data  assume $s=\bar s$.
\vspace{.5cm}

%%%%%%%%%%%%%%%%%%%%%%%%%%%%%%%%%%%%%%%%%%%%%%%%%%%%%%%%%%%%%%%%%%%%%%%%%%%%%%%%%%%%%%%%%%%%%%%%%%
{\bf 4. $s-\bar s$ and $D_d^{K^+-K^-}$, LO}
%%%%%%%%%%%%%%%%%%%%%%%%%%%%%%%%%%%%%%%%%%%%%%%%%%%%%%%%%%%%%%%%%%%%%%%%%%%%%%%%%%%%%%%%%%%%%%%%%%
\vspace{.5cm}

 We consider  $(\tilde\sigma_p + \tilde\sigma_n)^{K^+-K^-}$ and
 $(\tilde\sigma_p - \tilde\sigma_n)^{K^+-K^-}$.  In LO we have:
 %the charged Kaon production in SIDIS is:
 \be
 \tilde\sigma_d^{K^+-K^-}=(\tilde\sigma_p + \tilde\sigma_n)^{K^+-K^-}=\frac{1}{9} [
(u_V+d_V)\,(4D_u+D_d)^{K^+-K^-}+2(s-\bar s)D_s^{K^+-K^-} ]
\ee
\be
 (\tilde\sigma_p - \tilde\sigma_n)^{K^+-K^-}=\frac{1}{9} [ (u_V-d_V)\,(4D_u-D_d)^{K^+-K^-}]
\ee
We define the following measurable quantities:
\be
R_+ (x,z)=\frac{\sigma_{d}^{K^+}-\sigma_{d}^{K^-}}{u_V+d_V}=
(4D_u+D_d)(z)\left[1+\frac{ (s-\bar s)}{2(u_V+d_V)}\left(\frac{D_s}{D_u}\right)^{K^+-K^-}(z)\right]
\label{R+}
\ee
and
\be
R_- (x,z)&=&\frac{(\sigma_{p}-\sigma_{n})^{K^+-K^-}}{u_V-d_V}=(4\,D_u-D_d)^{K^+-K^-}(z)
\label{R-}
\ee
Note that the $x$-dependence in (\ref{R+}) is induced solely by the difference $s-\bar s$,
while in $R_-$ there is no $x$-dependence in LO. This result is independent of the FF.
 Then examining the $x$-dependence of $R_{\pm}(x,z_0)$ at some $z_0$,   we can deduce
the following:

1) if in some $x$-interval $R_+(x,z_0)$ is independent on $x$  then,
we can conclude that $(s-\bar s)=0$ in this $x$-interval.
Recall that since $D_s^{K^+ - K^-} $ is a favoured transition
$(D_s/D_u)^{K^+-K^-}> 1$.

2) if  $R_-(x,z_0)$ is also independent of $x$ , then  this suggests
that the LO approximation is reasonable.

 3) if $R_+(x,z_0)$ and $R_-(x,z_0)$ are \textit{both} independent of $x$, and if in addition,
 $R_+(x,z_0) = R_-(x,z_0)$, then both $s-\bar s=0$ in the considered $x$-interval
{\it and} $D_d^{K^+-K^-}(z_0)= 0$.

4) if $R_+(x,z_0)$ and $R_-(x,z_0)$ are \textit{both} independent of $x$,
 but they are \textit{not} equal, $R_+(x,z_0) \neq R_-(x,z_0)$, we conclude that $s-\bar
s=0$ in the considered $x$-interval, {\it but} $D_d^{K^+-K^-}(z_0)\neq 0$.

The above results 1) -- 4) are independent of our knowledge of the FFs.

5) if $D_d^{K^\pm}$ are known at some $z_0$, limits on $s-\bar s$ can be obtained. We have:
\be
\vert \frac{(s-\bar s)}{2(u_V+d_V)}
\left(\frac{D_s}{D_u}\right)^{K^+-K^-}\hspace{-0.8cm}(z_0)\vert\leq \frac{\delta r_+}{\vert r_+\vert }
\ee
where $\delta r_+/ r_+$ is the precision of the measurement:
$R_+(x,z_0)=r_+(z_0) \pm \delta r_+(z_0)$.

6) if $R_-(x,z)$ is not a function of $z$ only, then NLO corrections
are needed, which we  consider below.

The above tests for $s-\bar s =0$ and $D_d^{K^+-K^-}= 0$ can be
spoilt either by $s-\bar s \neq 0$ and/or $D_d^{K^+-K^-}\neq 0$,
or by NLO corrections, which are both complementary in size.
That's why it is important to formulate tests sensitive to $s-\bar
s =0$ and/or $D_d^{K^+-K^-}= 0$ solely, i.e. to consider NLO.
\vspace{0.5cm}

%%%%%%%%%%%%%%%%%%%%%%%%%%%%%%%%%%%%%%%%%%%%%%%%%%%%%%%%%%%%%%%%%%%%%%%%%%%%%%%%%%%%%%%%%%%%%%%%%%
{\bf 5. $s-\bar s$ and $D_d^{K^+-K^-}$, NLO}
%%%%%%%%%%%%%%%%%%%%%%%%%%%%%%%%%%%%%%%%%%%%%%%%%%%%%%%%%%%%%%%%%%%%%%%%%%%%%%%%%%%%%%%%%%%%%%%%%%
\vspace{0.5cm}

If an NLO treatment is necessary it is still possible to reach
some conclusions, though less detailed than in the LO case. We now
have:
 \be
 \tilde\sigma_d^{K^+-K^-}=
 \frac{1}{9} \left[(u_V+d_V)\otimes (4D_u+D_d)^{K^+-K^-}+2(s-\bar s)\otimes D_s^{K^+-K^-}\right]
 \otimes (1+\alpha_s\,C_{qq})\label{NLO+}
 \ee
 \be
(\tilde\sigma_p - \tilde\sigma_n)^{K^+-K^-}&=& \frac{1}{9} (u_V-d_V)
\otimes (1+\alpha_s\,C_{qq})\otimes (4D_u-D_d)^{K^+-K^-}\label{NLO-}
\ee

If instead of using (\ref{NLO+}) and (\ref{NLO-}),  we succeed to obtain
an acceptable fit for the $x$ and $z$-dependence of
both $p-n$ and $p+n$ data with the same fragmentation function $D(z)$:
\be
&&
(\tilde\sigma_p - \tilde\sigma_n)^{K^+-K^-}\approx \frac{4}{9} (u_V-d_V)\,\otimes\,(1+\alpha_s\,{\cal C}_{qq})\otimes\,D(z),\\
&&(\tilde\sigma_p +
\tilde\sigma_n)^{K^+-K^-}\approx \frac{4}{9}
(u_V+d_V)\,\,\otimes\,(1+\alpha_s\,{\cal C}_{qq})\otimes\,D(z).
\ee
than we can conclude that both $s-\bar s
\approx 0$ {\it and} $D_d^{K^+-K^-}\approx 0$, and that $D(z) =
D_u^{K^+-K^-}$.

Note that for all above tests, both in LO and NLO approximation,
we don't require a knowledge of $D_q^{K^+-K^-}$.
\vspace{0.5cm}

%%%%%%%%%%%%%%%%%%%%%%%%%%%%%%%%%%%%%%%%%%%%%%%%%%%%%%%%%%%%%%%%%%%%%%%%%%%%%%%%%%%%%%%%%%%%%%%%%%
{\bf 6. $\Delta s-\Delta\bar s$ in  $K^\pm$ production in SIDIS}
%%%%%%%%%%%%%%%%%%%%%%%%%%%%%%%%%%%%%%%%%%%%%%%%%%%%%%%%%%%%%%%%%%%%%%%%%%%%%%%%%%%%%%%%%%%%%%%%%%
\vspace{0.5cm}

Recently the COMPASS collaboration measured~\cite{COMPASS} the difference
asymmetry in SIDIS with longitudinally polarized muons and protons:
\be
A_d^{h-\bar h}=\frac{\Delta \tilde\sigma^{h-\bar h}}{\tilde\sigma^{h-\bar h}}.
\ee
and singled out the polarized valence quarks. Here we draw attention that
if the same asymmetry is measured
with final Kaons, information on $\Delta s-\Delta\bar s$ can be obtained:
\be
A_d^{K^+-K^-}(x,z)
%&=&\frac{\Delta \tilde\sigma^{K^+-K^-}}{\tilde\sigma^{K^+-K^-}}\\
\simeq\frac{\Delta u_V+\Delta d_V}{ u_V+ d_V}
\left\{1\!+\!\left(\!\frac{\Delta s-\Delta\bar s}{\Delta u_V+\Delta d_V} -
\frac{ s-\bar s}{u_V+ d_V}\right)\!\left(\frac{D_s}{2D_u}\right)^{K^+-K^-}\right\}
\ee
The $z$-dependence of $A_d^{K^+-K^-}$ is present only if $\Delta s-\Delta\bar s$ and/or
$ s-\bar s$ are non-zero. Thus, studying the $z$-dependence of $A_d^{K^+-K^-}$
one can obtain information about $\Delta s-\Delta\bar s\simeq 0$, suppose we already
have  the information about $ s-\bar s\simeq 0$, as discussed above.
%%%%%%%%%%
\vspace{0.5cm}

At the end a few remarks on the measurability of the discussed asymmetries. In general,
 these are difference asymmetries and high precision measurements are required.
 In addition, the data should be presented in bins in both $x$ and $z$.
Quite recently such binning was done in \cite{HERMES}
for the very precise data of the HERMES collaboration on $K^\pm$-production in SIDIS
  on proton and deuterium.
  These  results  show that for $0,350\leq z\leq 0,450$ and for $0,450 \leq z\leq 0,600$
  in the $x$-interval $0,023 \leq x\leq 0,300$
 the accuracy of the data allows to form the differences $(\sigma_d)^{K^+-K^-}$ and
 $(\sigma_p-\sigma_n)^{K^+-K^-}$ with errors not bigger than ~  7-13\% and ~10-15\% respectively.
 Then one can form the ratios
 $R_+$ and $R_-$ with these precisions and perform the above tests.
\vspace{0.5cm}

{\bf Acknowledgments}\\
 When this work was finished we understood about~\cite{Soffer},
 where similar questions were treated.
 This work was supported by a JINR(Dubna)-Bulgaria Collaborative Grant.

\end{document}